# Re-infection by untreated partners of people treated for *Chlamydia trachomatis* and *Neisseria gonorrhoeae*: mathematical modelling study

Nicola Low,[1] Janneke Cornelia Maria Heijne,[1,2] Sereina Annik Herzog,[1,3] Christian Lorenz Althaus[1]

[1] Institute of Social and Preventive Medicine (ISPM), University of Bern, Bern, Switzerland

[2] National Institute for Public Health and the Environment (RIVM), Bilthoven, The Netherlands

[3] Institute for Medical Informatics, Statistics and Documentation, Medical University of Graz, Graz, Austria

**Correspondence to** Christian Lorenz Althaus, Institute of Social and Preventive Medicine (ISPM), University of Bern, Bern, Switzerland; christian.althaus@alumni.ethz.ch; +41 31 631 56 40 (phone); +41 31 631 35 20 (fax)

# ABSTRACT

**Objectives:** Re-infection after treatment for *Chlamydia trachomatis* or *Neisseria gonorrhoeae* reduces the effect of control interventions. We explored the impact of delays in partner treatment on the expected probability of re-infection of index cases using a mathematical model.

**Methods:** We used previously reported parameter distributions to calculate the probability that index cases would be re-infected by their untreated partners. We then assumed different delays between index case and partner treatment to calculate the probabilities of re-infection.

**Results:** In the absence of partner treatment, the medians of the expected re-infection probabilities are 19.4% (interquartile range (IQR) 9.2–31.6%) for chlamydia and 12.5% (IQR 5.6–22.2%) for gonorrhoea. If all current partners receive treatment three days after the index case, the expected re-infection probabilities are 4.2% (IQR 2.1–6.9%) for chlamydia and 5.5% (IQR 2.6–9.5%) for gonorrhoea.

**Conclusions:** Quicker partner referral and treatment can substantially reduce re-infection rates for chlamydia and gonorrhoea. The formula we used to calculate re-infection rates can be used to inform the design of randomised controlled trials of novel partner notification technologies like accelerated partner therapy.

# KEY MESSAGES BOX

- Our formula calculates the probability of re-infection of index cases with chlamydia and gonorrhoea by their untreated partners.

- Quicker treatment of partners can substantially reduce re-infection rates for chlamydia and gonorrhoea.

- The intuitive and flexible formula can be easily adapted to inform the design of studies such as randomised controlled trials.

# INTRODUCTION

If sexual partners are not treated concurrently for *Chlamydia trachomatis* (chlamydia) or *Neisseria gonorrhoeae* (gonorrhoea), the index case may be re-infected.[1 2] Our previous work has shown that the impact of screening interventions for chlamydia infection is limited by re-infection from untreated partners.[3] Repeated infection might also increase the risk of adverse sequelae, particularly in women.

Partner notification (PN) limits re-infection in people who have been treated for chlamydia or gonorrhoea. Expedited partner therapy and accelerated partner therapy are PN technologies that reduce the time between diagnosis and treatment of the index patient to partner referral and treatment.[1 4] In randomised controlled trials (RCT), expedited partner therapy reduced the incidence of repeated detection of either chlamydia or gonorrhoea infection.[1] It was not possible to determine the proportion of repeated infections that resulted from re-infection by an untreated partner, infection from a new partner, or antibiotic treatment failure. The incremental benefits of reducing the delay to partner treatment on re-infection of index cases by partners who have not yet been treated are not well understood.

The effects of PN can be explored with mathematical models. Dynamic transmission models often assume that index cases and their partners are treated at the same time so the risk of re-infection with different delays to partner treatment cannot be studied.[3] Calculating the probability of re-infection requires a model that takes into account competing risks of re-infection, partner treatment, clearance of infection in the partner, and dissolution of the sexual partnership. Such a model could help us better understand the differences between chlamydia and gonorrhoea re-infection after treatment.

The objective of this study was to develop a model to calculate the expected probability with which treated index cases will be re-infected with chlamydia and gonorrhoea by their untreated partners. We explore the effect that varying the time delay to partner treatment has on re-infection of index cases.

## METHODS

## Probability of re-infection

We devised a formula to calculate the probability of re-infection of index cases by their as yet untreated partners:

$$P = p_p \times p_i \left( \frac{f \times \beta}{f \times \beta + \gamma + \sigma + \delta} + (1 - \varepsilon) \times \frac{\delta}{f \times \beta + \gamma + \sigma + \delta} \times \frac{f \times \beta}{f \times \beta + \gamma + \sigma} \right)$$

$p_p$ is the probability that the index case is in an ongoing sexual partnership and $p_i$ is the probability that the partner is infected. The first term in the brackets represents the probability that the index case is re-infected before the partner is treated, the partner's infection is cleared, or the sexual partnership is dissolved. The successfully treated index case is assumed to engage in new sex acts with the partner at a frequency $f$, where the infection can transmit at a per sex act transmission probability $\beta$. Transmission can occur until the partner is either treated at rate $\delta$ ($1/\delta$ is the average delay in the time to partner treatment), the infection is spontaneously cleared in the partner at rate $\gamma$, or the sexual partnership dissolves at rate $\sigma$. The second term in the brackets corresponds to the probability that re-infection occurs in cases in which the partner has not been successfully treated, where $\varepsilon$ is the treatment efficacy in the partner. We assume that all probabilities and events are independent of each other.

## Parameter values

To calculate the expected range of re-infection probabilities, we generated $10^6$ parameter sets by randomly drawing from uniform ($U(a, b)$ from $a$ to $b$) or binomial ($B(n, p)$ with sample size $n$ and probability $p$) distributions. Unless otherwise noted, the parameter ranges are based on the studies by Althaus et al.[5] and Garnett et al.[6] and the papers they cite. The probability that index cases are in an on-going partnership was assumed to be in the interval $U(0, 1)$. Chlamydia or gonorrhoea positivity in current partners is drawn from $B(77, 0.69)$ and $B(188, 0.80)$.[7] We assumed an average duration of sexual partnerships in chlamydia-infected individuals from $U(1 \text{ week}, 6 \text{ months})$. For gonorrhoea-infected individuals, we assumed that sexual partnerships are shorter ($U(1 \text{ day}, 2 \text{ weeks})$). For both infections, the frequency of sex acts was drawn from $U(1, 7)$ per week.[8] The per sex act transmission probabilities for chlamydia and gonorrhoea were drawn from $U(0.06, 0.167)$ and $U(0.19, 0.53)$. The average duration of chlamydia and gonorrhoea infection in the current partner is drawn from $U(6, 12)$ and $U(1, 6)$ months. We assume that the PN rate is 100%, and that the treatment efficacy for a notified partner was 90% for chlamydia and 100% for gonorrhoea.

## RESULTS

We compare re-infection with chlamydia and gonorrhoea in the absence of PN to expected re-infection in the presence of PN and used the following average delays to treatment of the current partner: 14 days, 3 days, and 1 day (figure 1). Without PN, median re-infection probabilities from the $10^6$ parameter combinations are 19.4% (interquartile range (IQR) 9.2-31.6%) for chlamydia and and 12.5% (IQR 5.6–22.2%) for gonorrhoea. The values for chlamydia re-infection support previous observations in empirical and mathematical modeling studies.[2][9] Less is known about gonorrhoea re-infection in the absence of PN. Although the per sex act transmission probability for gonorrhoea is higher than for chlamydia, we find a lower probability of re-infection because sexual partner change is more frequent and partnerships do not last as long in gonorrhoea-infected individuals.

Reducing the delay between index case and partner treatment from 14 days[10] to 1 or 3 days[4] substantially reduces the risk of re-infection. For chlamydia, median probabilities of re-infection are 8.8% (IQR 4.3–14.6%), 4.2% (IQR 2.1–6.9%) and 2.8% (IQR 1.4-4.6%) for decreasing delays of partner treatment. For gonorrhoea, the respective probabilities are 9.7% (IQR 4.5–17.1%), 5.5% (IQR 2.6-9.5%) and 2.6% (IQR 1.3-4.6%). Thus, the probability of re-infection after treatment for chlamydia and gonorrhoea become more alike as the time to partner treatment grows shorter.

## DISCUSSION

Our formula provides a general framework within which one can assess the risk that index cases treated for chlamydia or gonorrhoea will be re-infected by their, as yet, untreated partners. We found that reducing the time to partner treatment to 1 to 3 days can reduce re-infection of treated index cases from untreated partners substantially. The formula is intuitive and flexible, and can be easily adapted to model variations in the mechanisms of transmission and PN. For example, it can easily accommodate reduced frequency of sexual intercourse or condom use.

Because we do not know all the parameters that influence transmission of chlamydia and gonorrhoea, the generalisability of our findings is limited. There are more studies with prospectively collected data about chlamydia re-infection than for gonorrhoea, at least for women.[2] Re-infection rates might differ between women and men but we did not consider sex-specific parameters in this study. We had information about the average length of sexual partnerships of chlamydia-infected[5] but not gonorrhoea-infected individuals. We assumed that treated index cases can only be re-infected by a single current partner but acknowledge that concurrent sexual partnerships also exist. Finally, our study concentrates only on re-infection of index cases and does not investigate the effects that different periods of delay to partner treatment might have on transmission in the population. Without dynamic transmission models we could not address the latter question, but we have shown recently that preventing chlamydia re-infection of the treated index is the most effective way of increasing the population-level effect of PN.[3]

The results from our formula can be compared with findings from empirical studies. For example, recurrent or persistent chlamydia and gonorrhoea infection were measured in an RCT of the effects of expedited partner therapy.[1] Our calculated risk of re-infection for a PN delay of 14 days are consistent with the RCT findings for standard referral of partners. Golden et al.[1] further found that expedited partner therapy reduced persistent or recurrent infections more for gonorrhoea than for chlamydia. Our study did not replicate this finding but this might be because one could not distinguish between treatment failure in index cases and repeated infection from current or new partners in the RCT.

Our model can help to understand the potential impact on chlamydia and gonorrhoea re-infection of reducing delays to partner treatment. Its value will increase as we improve our estimates of transmission and sexual behaviour, particularly for individuals infected with gonorrhoea. We conclude that PN technologies that shorten delays to partner treatment, such as expedited and accelerated partner therapy, substantially reduce index case re-infection by untreated partners, when compared with standard patient referral.

## Funding

This study is part of a project funded by the NIHR Health Technology Assessment programme (project number 07/42/02). The project will be published in full in the Health Technology Assessment journal series. Visit the HTA programme website for more details (www.hta.ac.uk/1722). The views and opinions expressed therein are those of the authors and do not necessarily reflect those of the Department of Health. CLA, JCMH and SAH acknowledge financial support by the Swiss National Science Foundation (project numbers PZ00P3_136737, 320030_135654, PDFMP3_124952).

## Competing interests

In 2010, JCMH and NL received fees from GlaxoSmithKline for attending a meeting about chlamydia vaccines.

# Figure legends

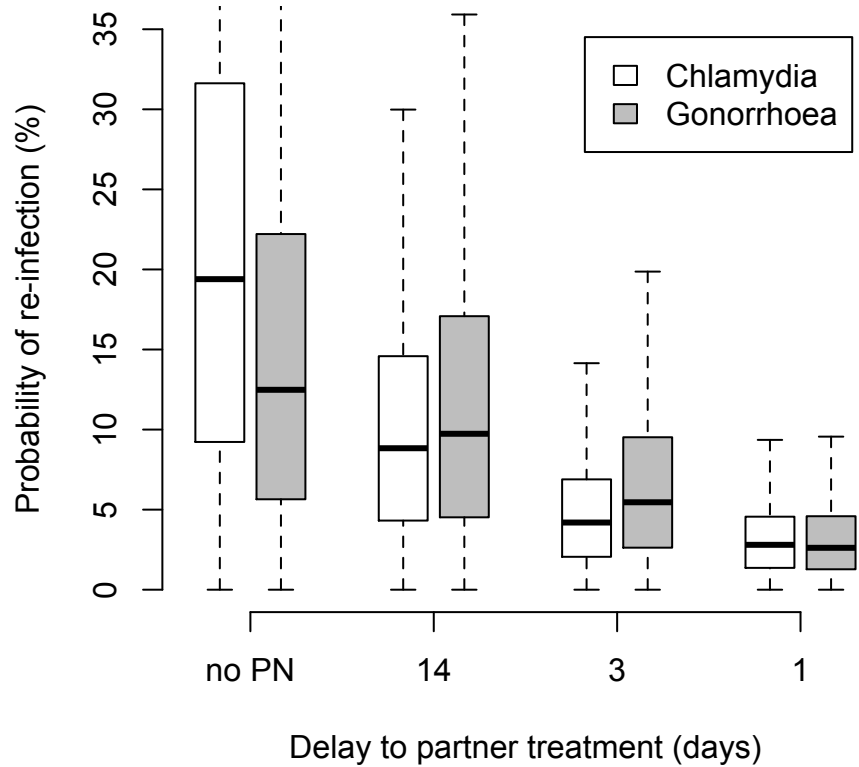

**Figure 1** Probabilities of re-infection of treated index cases by their untreated partners. Different delays to partner treatment are contrasted to the scenario without any partner notification (no PN). The figures show box-plots of $10^6$ different parameter combinations sampled from the distributions given in *Methods*. The whisker of the first two box plots extend to 65.2% for chlamydia and 47.1% for gonorrhoea.